# PERIOD VARIATIONS AND POSSIBLE THIRD COMPONENTS IN THE ECLIPSING BINARIES AH TAURI AND ZZ CASSIOPEIAE

D.E. Tvardovskyi[1,2], V.I. Marsakova[1,2], I.L. Andronov[2], L.S. Shakun[3]

[1] Odessa I. I. Mechnikov National University, Odessa, Ukraine,

*dmytro.tvardovskyi@gmail.com, vmarsakova@onu.edu.ua*

[2] Department "Mathematics, Physics and Astronomy", Odessa National Maritime University

Odessa, Ukraine, *tt_ari@ukr.net*

[3] Astronomical Observatory, Odessa I. I. Mechnikov National University, Odessa, Ukraine, *leomspace@gmail.com*

**Abstract:** In our research, we investigated two variable stars: AH Tau and ZZ Cas. They are eclipsing binary stars of W Ursae Majories and β Lyrae types. The period between eclipses of these stars changes with time. The reason for steady changes of the period could be the mass transfer (the flow of matter) between components of these stellar systems. For ZZ Cas the changes of the period are cyclic. That is why we assume the cyclic changes could be caused by the presence of the hypothetical third component (either a small star or a large planet). The cyclic changes of the period for AH Tau superimpose on steady ones (the period decrease). Thus we assume the third component and mass transfer are present. We also assume that the third components do not take part in the eclipses. However, due to their gravity, they make the visible close binary systems rotate and become closer or further to an observer. This explanation is called Light-Time Effect (LTE). Generally, an orbit of a third component is not a circle, but an ellipse and it is inclined relatively to the observer's line of sight. Using special plot called O-C curve we estimated the parameters of a third component's orbit such as a semi-major axis, an eccentricity, angles of orientation and a period of a third component's rotation. The O-C curve is the plot which shows how the difference between an observed and calculated moment of minimal brightness changes during a long period of time (usually it is several decades). To do this we created a modeling computer program using the computer language Python. In addition, we can calculate errors of third component's orbit parameters and even estimate its mass. The values of masses of the third components within errors of calculations show that the third components are probably stars. All these calculations were made using all available data from international database BRNO (Brno Regional Network for Observers). Moreover, we used moments of minima which we calculated as the result of observation processing from AAVSO database (American Association of Variable Stars Observers). These results are provided in the tables and plots.

**Абстракт:** У нашій роботі було досліджено дві змінні зорі: AH Tau та ZZ Cas. Вони є затемнювано-подвійними зорями типів W Великої Ведмедиці та β Ліри. Період між затемненнями цих зір змінюється з часом. Причиною постійних змін періоду може бути перетікання речовини (потік речовини) між зорями. У ZZ Cas зміни періоду є циклічними. Тому ми припускаємо, що вони можуть бути викликані наявністю гіпотетичного третього компонента (малої зорі або великої планети). У AH Tau циклічні зміни періоду накладаються на постійні (період зменшується). Отже ми припускаємо, що присутні й перетікання речовини, й третій компонент. Ми також припускаємо, що треті компоненти не беруть участі в затемненнях, але завдяки своїй силі тяжіння вони змушують видимі тісні подвійні системи наближатися або віддалятися від спостерігача. Це пояснення називається ефектом затримки світлового сигналу (LTE). У загальному випадку орбіта третього компонента не є колом, а еліпсом і нахилена відносно променю зору спостерігача. Використовуючи спеціальний графік, який називається кривою О-С, ми оцінюємо такі параметри орбіти третього компонента, як велику піввісь, ексцентриситет, кути орієнтації та період обертання третього компонента. Крива О-С – це графік, який показує, як різниця між спостережуваним і розрахованим моментами мінімальної яскравості змінюється протягом тривалого проміжку часу (як правило, це кілька десятиліть). Для цього ми створили комп'ютерну програму для моделювання, використовуючи мову програмування Python. Крім того, ми можемо обчислити похибки параметрів орбіти третього компонента і навіть оцінити його масу. Маса третіх компонентів в межах похибок розрахунків показує, що треті компоненти, напевно, є зорями. Всі ці розрахунки були зроблені з використанням усіх наявних даних з міжнародної бази даних BRNO (Brno Regional Network for Observers) та моментів мінімумів, які ми розрахували з використанням спостережень з бази даних AAVSO (American Association of Variable Stars Observers). Ці результати наводяться у таблицях та на графіках.

## Introduction

For this research, we chose two eclipsing variable stars with period changes that are cyclic, but not sinusoidal: AH Tau and ZZ Cas. The analysis for the several eclipsing variables with cyclic period changes we discussed in our previous papers (Tvardovskyi; Marsakova & Andronov (2017) and Tvardovskyi & Marsakova (2015)). The main observational parameters of the variables were taken from General Catalogue of Variable Stars (GCVS) and listed in Table 1. Some fundamental parameters were published in articles Xiang et al. (2015) and Liao & Qian (2009). Geometrical and physical parameters of the stellar systems were computed there. In the current article we compare parameters of the third components such as masses and orbital elements with ones published in the previous articles of other researchers.

Moreover, there is some additional information about these stars as the result of previous researches made by other astronomers. For both stellar systems, the presence of the third components was suspected by other authors, but for ZZ Cas even presence of the forth component was supposed. Mass transfer/loss was supposed for AH Tau. For both stars, geometrical and physical parameters were computed.

This study was carried out in a frame of the projects "Inter-Longitude Astronomy" (Andronov et al., 2017) and "AstroInformatics" (Vavilova et al., 2017) projects.

Table 1. Observational parameters provided in General Catalogue of Variable Stars (GCVS, Samus et al., 2018) and by Xiang et al. (2015) and Liao & Qian (2009).

| Parameter | AH Tau | ZZ Cas |
|---|---|---|
| RA | $03^h47^m11.98^s$ | $00^h33^m30.38^s$ |
| DEC | 25°06'59.5" | 62°30'40.2" |
| Type of variability | EW/KW | EB |
| Initial epoch (JD) | 31062.5081 | 33437.5 |
| Period (days) | 0.3326754 | 1.243527 |
| Spectral type | G1p | B3 |
| $M_1$, $M_\odot$ | 1.04 | 7.6 |
| $M_2$, $M_\odot$ | 0.52 | 5.32 |
| Article | Xiang et al. (2015) | Liao & Qian (2009) |

Table 2. Articles of other authors where different types of results were provided. The first column identifies the types of information, which is necessary for the current research. Next two columns are the lists of references for articles where these kinds of calculations were published.

| | AH Tau | ZZ Cas |
|---|---|---|
| Ephemerid correction | - | Taylor (1941); Frieboes-Conde & Herczeg, (1971); Kreiner & Tremko, (1991). |
| Spectral observations | - | Struve (1947); Taylor (1941). |
| Presence of a third component | Xiang et al. (2015); Yang et al (2010); Zasche et al (2008); Lee et al (2004). | Frieboes-Conde & Herczeg (1971); Kreiner & Tremko (1991); Liao & Qian (2009). |
| Presence of a 4th component | - | Liao & Qian (2009). |
| Mass transfer/loss | Xiang et al. (2015); Yang et al (2010); Zasche et al (2008); Lee et al (2004); Yulan Yang & Qingyao Liu (2002). | - |
| Parameters of the system | Xiang et al. (2015); El-Sadek et al (2009); Yang et al (2010); Byboth et al (2004); Liu et al (1991). | Struve (1947); Liao & Qian (2009). |

### O-C analysis

To build O-C curves we used observations from different sources. Most of them were taken from Brno Regional Network of Observers (BRNO). In addition, we computed moments of minima for all available data from American Association of Variable Stars Observers database (AAVSO) using the software MAVKA (Andrych et al., 2017). For all available moments of minima, we calculated the values of O−C and built the O−C curve. In our research, we investigated stellar systems, which have one of two different types of O-C curves:

- cyclic period changes (they correspond to possible presence of the third component);
- superposition of cyclic period changes and parabolic trend (they correspond to both presence of the third component and mass transfer).

More details about possible explanations of different types of period changes are provided in Tvardovskyi, Marsakova & Andronov (2017) and Tvardovskyi & Marsakova (2015).

### Orbital elements of the third body

In both stars we detected that cyclic period changes on O-C curve are similar to sinusoidal, but they have clear asymmetry. We suppose that this asymmetry is caused by the elliptical shape and inclination of the third bodies' orbits. We assume that the visible components of the stellar systems are close to each other and the third one has a lot wider elliptical orbit. Each orbit we can describe by its seven parameters, such as:

- semi-major axis (a);
- eccentricity (e);
- three angles of inclination (argument of the pericenter ($\Omega$), inclination (i) and longitude of pericenter ($\omega$));
- orbital period (T);
- moment of the pericenter transit ($t_0$).

We decided to calculate these parameters and create a model O-C curve in the case of the presence of the third body in the system. To make the calculations we wrote a computer program on the computer language Python 3.5. This program uses the Levenberg-Marquard algorithm (Marquardt, 1963, Moré, 1977), which represents the least squares method for the non-linear model. This algorithm builds the model O-C curve for initial parameters (this is our own function), then corrects them and repeats the procedure until the squares of the differences will be as little as possible.

Using the Kepler's equation and formulas of the coordinate system turn, we can easily write the equation which describes the shape of O-C curve for different orbital elements:

$$\Delta t = \frac{a}{c}\sin i \left((\cos E - e)\sin \omega + \sin E \cos \omega \sqrt{1-e^2}\right) \quad (1)$$

As we can see from this formula, we cannot calculate all the orbital elements. Particularly, we cannot compute the argument of the pericenter or the major semi-axis and inclinational angle separately. We can calculate only the projection of the semi-major axis on the line of sight. Moreover, AH Tau has a superposition of a parabolic trend and cyclic period changes. For this star, we modified Eq. (1) to consider the parabolic trend:

$$\Delta t = \alpha t^2 + \beta t + \gamma + \frac{a}{c}\sin i \cdot \\ \cdot \left((\cos E - e)\sin \omega + \sin E \cos \omega \sqrt{1-e^2}\right) \quad (2)$$

Here $\alpha$, $\beta$ and $\gamma$ are the coefficients of the parabolic trend.

Before making calculations, our program needs the list of initial values of the parameters (Table 4). We estimated the amplitude and period of the O-C cyclic changes 'by eye'. We set the argument of the pericenter as $\pi$ radians or 180 degrees, as the average of the least and the largest possible. Eccentricity was set as 0.1. Coefficients $\alpha$, $\beta$, $\gamma$ and $t_0$ are set as equal to zero.

After calculations, the program returns the list of corrected elements with their errors (Table 2) and the picture (Fig. 2-5).

This is the table with parameters of the binary system orbit about the common barycenter. However, most of them will be the same for the third component's orbit. These ones are eccentricity, pericenter argument, initial moment of minima and rotation period. Semi-major axis of the third body's orbit will be the same number times bigger than one in the binary system, than relation of the masses of the binary system and of the third component.

### Taking into consideration the weights

In the investigated systems, we detected some systematic deviations between model O-C and observations. Some of them are caused by the large scatter of old visual observations. It happens if we assume that all observations have the same "importance" for approximation. Thus, we decided to analyze which types of observations exist on O-C curves. There are 4 main types of observations: visual ("vis"), photometric ("ccd"), photoelectric ("pe") and ones which were made using photographic plates ("pg", that are similar to visual observations).

Using this information, we created the system of weights. Weight of the point is a multiplier of the deviation between observations and model value of O-C for this point in the sum of the squares of deviations. It allows to take in consideration different accuracies of the minimum brightness determination. The better is the accuracy of the O-C value determination – the larger is the weight of the point. For "vis" and "pg" observations we set the weight equal to 1. For all other types of observations ("ccd", "pe") we set the weight equal to 10. This ratio (1 to 10) is equivalent to difference by one order in the accuracies of moments of minimal brightness determination according to the number of significant digits given in BRNO database. The visual observations were provided there with 3 digits after comma, and the photometric ones – with 4 digits, that corresponds the difference in accuracy of 10 times.

### Ephemerid correction

Ephemerid is the special formula for calculation the moment of minima:
$$T_c = T_0 + P \cdot E \quad (3)$$
The third summand appears in the case of parabolic trend presence:
$$T_c = T_0 + P \cdot E + P' \cdot E^2 = T_0 + P \cdot E + \dot{P} \cdot P \cdot E^2 \quad (4)$$
Here P is a period; E is a cycle number; $T_0$ is a moment of minima, which we set as the zero point and called the initial epoch; $P' = P \cdot \dot{P}$ is the rate of the period changes (days per cycle).

We used coefficients of approximation $(\alpha, \beta, \gamma)$ for correction.
- Coefficient $\alpha$ is the error of the value of the initial epoch;
- Coefficient $\beta$ is the error of the value of a period;
- Coefficient $\gamma$ is the quadratic addition, which describes the rate of the period changes (see section "Rate of the mass transfer").

Such correction was done for the stellar system AH Tau. The corrected ephemerid is:

$$T_c = (31062.7754 \pm 0.010) + (0.33266942 \pm 0.00000014) \cdot E + (-9.28 \pm 0.15) \cdot 10^{-11} \cdot E^2$$

### Masses of the third bodies

To estimate the masses of the third components, we suppose that in both researched systems the distance between the third component and two visible ones is much larger than the distance between visible components. In this case we used third Kepler law and the formula for the barycenter position. We derived the minimal possible mass of the third component after simplifications:

$$M_3 = \frac{c \Delta t}{\sqrt[3]{G}} \left[ \frac{2\pi}{T} \cdot (M_1 + M_2 + M_3) \right]^{\frac{2}{3}} \quad (5)$$

Here:
- G is the gravitational constant;
- T is the orbital period of the third body (and the period of O-C changes);
- $M_1$, $M_2$, $M_3$ are the masses of the components;
- c is the speed of light;
- $\Delta t$ is semi-amplitude of O-C cyclic changes.

After several iterations we got mass of the third component. In addition, we estimated errors of calculations using formula (6):

$$\sigma M_3 = \left( \frac{1}{M_3} - \frac{2}{3M} \right)^{-1} \cdot \sqrt{ \frac{\sigma_{\Delta t}^2}{\Delta t^2} + \frac{4}{9} \left[ \frac{\sigma_{M_1}^2}{M_1^2} + \frac{\sigma_{M_2}^2}{M_2^2} + \frac{\sigma_T^2}{T^2} \right] } \quad (6)$$

Here:
- $M = M_1 + M_2 + M_3$ – total mass of a triple system (visible components and a third one);
- $\sigma_{\Delta t}$ – error of semi-amplitude of O-C cyclic changes;
- $\sigma_{M_1}$ and $\sigma_{M_2}$ – errors of masses (they were not provided in the articles, so for both stars we estimated them as 5% of the masses);
- $\sigma_T$ – error of a period of cyclic changes.

More details about calculation of minimal mass of the third component are provided in Tvardovskyi, Marsakova & Andronov (2017).

### Rate of the mass transfer

If a star has a parabolic O−C curve or a cyclic one with parabolic trend, its period changes steadily. As the result of the O−C analysis, we detected that AH Tau has steady period changes, which we interpreted as the mass transfer from one of the components to another. To calculate the rate of the mass transfer, we used the formula (Andronov, 1991):

$$\dot{M} = \frac{1}{3} \frac{\dot{P}}{P} \frac{M_1 M_2}{(M_1 - M_2)} = \frac{1}{3} \frac{P'}{P^2} \frac{M_1 M_2}{(M_1 - M_2)} \quad (7)$$

In this formula: $\dot{M}$ is a rate of the mass transfer, $M_1, M_2$ are masses of the components in a close binary system, P is a period of variability, $\dot{P}$ is a derivative of the period with the time, $P' = 2\gamma$ is the derivative of the period with cycle number (Andronov, 1991). To obtain $\dot{M}$ in the standard units used in astronomy (solar masses per year) we must multiply it by the average quantity of days in the year (365.25).

Classical monographs on study of eclipsing binaries are e.g. by Tsesevich (1973) and Hofmeister et al. (1980).

### Conclusions and discussions

We studied two eclipsing binary stars with period changes: AH Tau and ZZ Cas. For both of these stars we calculated minimal possible mass of the third components. The mass of the third component for AH Tau equals to $0.37 M_\odot$, that corresponds to the parameters of an M-type star.

Our value is approximately 50% larger than value obtained by Dong Joo Lee et al (2004). For ZZ Cas the third component was supposed for the first time. According to its mass of about $3.9\ M_\odot$ (that is 20% smaller than the value published in Liao & Qian (2009)) the third component is rather a B-type star. Unfortunately, errors in these two articles were not published, so it is hard to say whose results are correct. For both stars we calculated parameters of the third bodies orbits. For AH Tau they are generally close to the ones published in Lee et al (2004). However, there are some deviations, especially in the value of the longitude of pericenter. In addition, we calculated the rate of the mass transfer for AH Tau. The similar calculation was done in Lee et al (2004). Nevertheless, there is 17 times difference between our and their results. Such a huge deviation could be explained by a small amount of observations and shorter time interval that is crucial for the accurate calculations.

**Acknowledgments:** We sincerely thank the AAVSO (Kavka, 2018) and BRNO associations of variable stars observers for their work that has made this research possible.

Table 3. Initial values of orbital elements of researched stars.

| Parameter | Units of measurement | I Boo | ZZ Cas |
|---|---|---|---|
| $\alpha$ | $10^{-12}$ days$^{-1}$ | 0 | - |
| $\beta$ | $10^{-7}$ | 0 | - |
| $\gamma$ | days | 0 | - |
| $a \cdot \sin i$ | $10^6$ km | 2000 | 2000 |
| e | 1 | 0.1 | 0.1 |
| $\omega$ | degrees | 180 | 180 |
| t0 | JD-240000, $10^3$ days | 5000 | 0 |
| T | $10^3$ days | 20000 | 7000 |
| Trend | | parabolic | none |

Table 4. Corrected values of orbital elements of researched stars.

| Parameter | Units of measurement | AH Tau | | ZZ Cas |
|---|---|---|---|---|
| $\alpha$ | $10^{-12}$ days$^{-1}$ | -276±4 | -164 | - |
| $\beta$ | $10^{-7}$ | 178±4 | - | - |
| $\gamma$ | $10^{-3}$ days | -266±10 | - | - |
| $a \cdot \sin i$ | $10^6$ km | 440±5 | 426 | 624±4 |
| e | 1 | 0.330±0.014 | 0.52 | 0.582±0.005 |
| $\omega$ | degrees | 112.8±3.0 | 164.4 | 192.9±0.2 |
| $t_0$ | JD-240000, $10^3$ days | 12.204±0.0.20 | 45.452 | 2.28±0.08 |
| T | $10^3$ days | 15.53±0.09 | 12.93 | 6.79±0.01 |
| $\sigma$ | days | 0.00544 | - | 0.00312 |
| Source | | This research | Dong Joo Lee et al, 2004 | This research |

Table 5. Minimal possible masses of the third components and rate of the mass transfer.

| | Rate of the mass transfer (Solar masses per year) | | Minimal mass of the third component (Solar masses) | |
|---|---|---|---|---|
| AH Tau | (6.3±0.7)·$10^{-7}$ | 3.8·$10^{-8}$ [Dong Joo Lee et al (2004)] | 0.37±0.02 | 0.24 [Dong Joo Lee et al (2004)] |
| ZZ Cas | - | - | 3.90±0.22 | 5.00 [Liao & Qian (2009)] |
| Source | This article | Other authors | This article | Other authors |

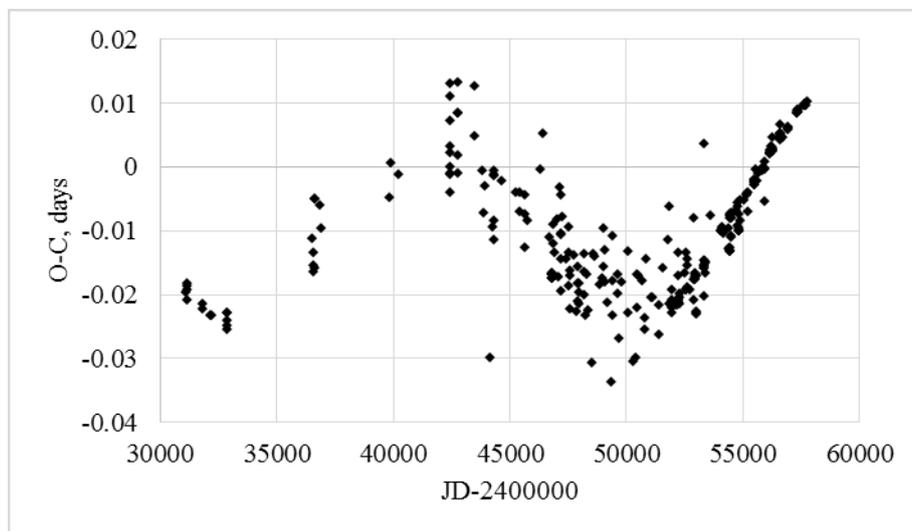

Fig. 1. O-C curve of AH Tau with subtracted parabolic trend.

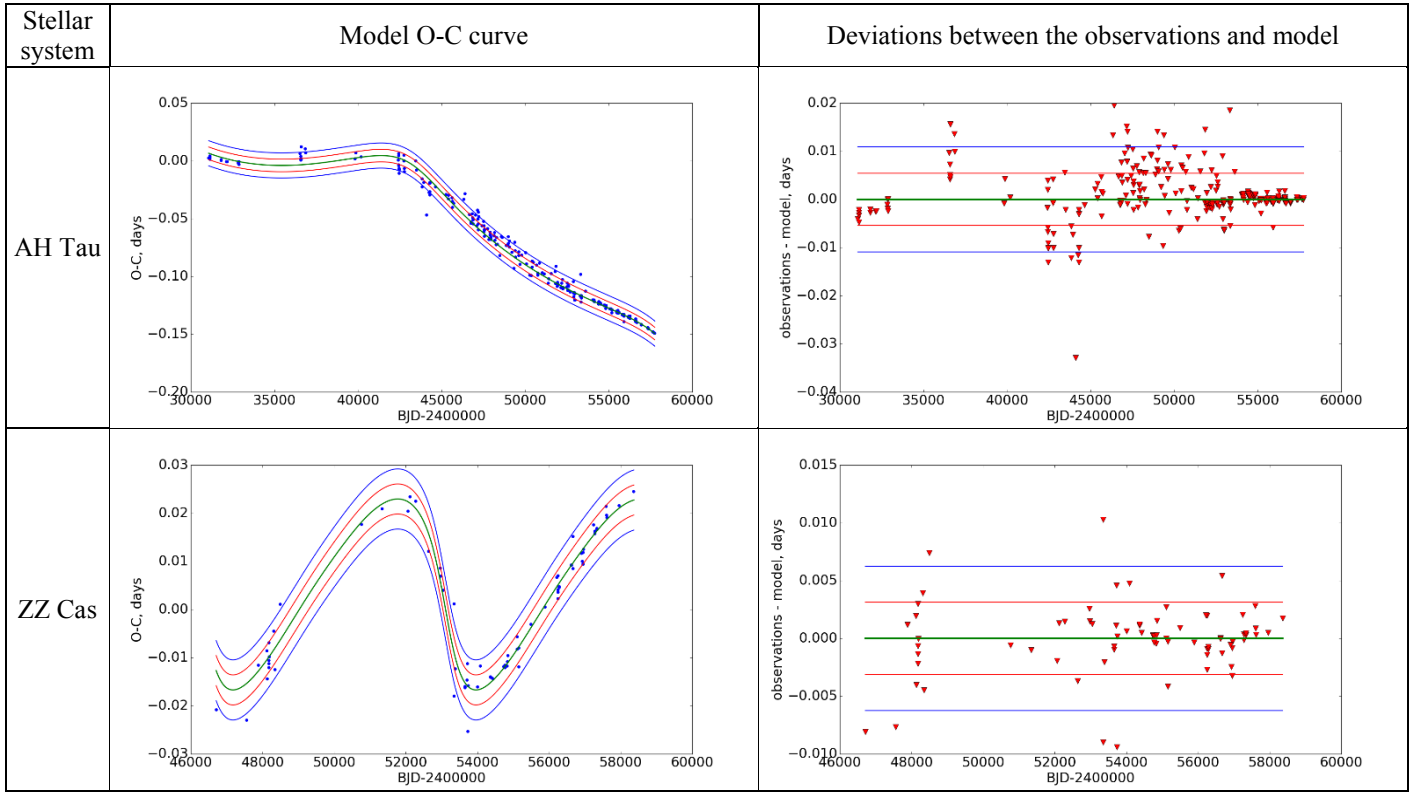

Fig. 2-5. Plots of approximation of O-C curve by model of the elliptical orbit and deviations between observational data and model O-C curve. In the second column, the dots are observations, the solid green line is the model O-C for corrected values of the orbital elements, red and blue lines correspond to 1σ and 2σ intervals.

Table 6. Obtained moments of minima using the time series observations from the AAVSO database.

| Observer code | Filter | BJD of the minimum | Error | Degree of polynomial |
|---|---|---|---|---|
| ZZ Cas | | | | |
| BM | Vis. | 56620.57802 | 0.00014 | 8 |
| SAH | V | 57279.65393 | 0.00016 | 8 |
| SNE | V | 57330.63960 | 0.00023 | 8 |
| SAH | V | 57611.67947 | 0.00027 | 8 |
| SAH | V | 58367.74877 | 0.00042 | 8 |
| AH Tau | | | | |
| DKS | V | 54399.89762 | 0.00011 | 8 |
| DKS | V | 54402.72580 | 0.00009 | 8 |
| DKS | V | 54402.89155 | 0.00008 | 8 |
| DKS | V | 54429.67196 | 0.00011 | 8 |
| GKA | V | 54475.58074 | 0.00009 | 6 |
| GKA | V | 54429.83824 | 0.00009 | 8 |
| GKA | V | 54512.67332 | 0.00015 | 8 |
| GKA | V | 54736.89537 | 0.00008 | 8 |
| GKA | V | 54829.54492 | 0.00013 | 8 |
| WAB | V | 54830.54276 | 0.00008 | 8 |
| WAB | V | 54837.52885 | 0.00009 | 8 |
| WAB | V | 54848.67374 | 0.00014 | 8 |
| WAB | V | 54399.89762 | 0.00011 | 8 |
| WAB | V | 54402.72580 | 0.00009 | 8 |